\documentstyle[preprint,aps]{revtex}

\begin{document}
\title{Solution of Dirac equation in the near horizon geometry of an extreme Kerr
black hole.}
\author{I.Sakalli\thanks{%
izzet.sakalli@emu.edu.tr } and M.Halilsoy\thanks{%
mustafa.halilsoy@emu.edu.tr}}
\address{Physics Dept. EMU G.Magosa, Mersin 10, Turkey.}
\date{\today}
\maketitle

\begin{abstract}
Dirac equation is solved in the near horizon limit geometry of an extreme
Kerr black hole. We decouple equations first as usual, into an axial and
angular part. The axial equation turns out to be independent of the mass and
is solved exactly. The angular equation reduces, in the massless case, to a
confluent Heun equation. In general for nonzero mass, the angular equation
is expressible at best, as a set of coupled first order differential
equations apt for numerical investigation. The axial potentials
corresponding to the associated Schr\"{o}dinger-type equations and their
conserved currents are found. Finally, based on our solution, we verify in a
similar way the absence of superradiance for Dirac particles in the near
horizon, a result which is well-known within the context of general Kerr
background.
\end{abstract}

\section{INTRODUCTION}

\bigskip During the last three decades the study of spin-$\frac{1}{2}$
particles on type-D spacetimes has attracted much interest and by now
accumulated results are already available in the literature [1--3] (and
references cited therein). The main reason for this is that all well-known
black holes (BHs) are in this category and their better understanding
involves detailed analysis of various physical fields in their vicinity.
Dirac particles with (and without) mass constitutes one such potential
candidate whose interaction/behavior around BHs may reveal information of
much significance. Test Dirac equation in spacetimes other than BHs are also
started to \ arouse interests for various reasons. From this token we cite
the Robertson-Walker and Bertotti-Robinson (BR) spacetimes [4--6]. The
latter in particular has already gained much recognition in connection with
extremal BHs, higher dimensions and the brane world.

In this paper we consider the Dirac equation in the near horizon geometry of
an extreme Kerr BH. This constitutes the most important region of the
outerworld prior to the horizon of a Kerr BH enhanced with the extremality
condition. Extremal BHs are believed to have connection with the ground
states of quantum gravity. This alone justifies, in spite of the absence of
backreaction effects, the importance of spin-$\frac{1}{2}$ particles on such
backgrounds. The throat geometry is a completely regular vacuum solution
with an enhanced symmetry group $SL(2,R)\times U(1)$. In many aspects, this
solution shares the common features with the $AdS_{2}\times S_{2}$ geometry
arising in the near horizon limit of extreme Reissner-Nordstr\"{o}m BH. The
behavior of massless scalar fields in the extreme Kerr throat has been
considered, and it is found that certain modes with large azimuthal quantum
number exhibits superradiance [7]. This implies that the geodesics near the
horizon can escape to infinity carrying energy-momentum more than the amount
that infalls. Our solution enables us to investigate a similar phenomenon
with Dirac fields which turns out to be negative as far as superradiance is
conserved. This result is in accord with the treatment of Dirac equation in
the general Kerr background in the absence of an exact solution [1]. The
analysis of fermions in Kerr-Newman background and absence of \
superradiance was shown first by Lee [3]. Our aim is to revisit this item --
not only by separating equations and deducing results on general grounds --
but rather obtaining exact solutions and employing them. The advantage of
confining ourselves to the near horizon alone bears its fruits by allowing
solutions expressible in terms of known polynomials. In the general
background of Kerr family of BHs even this much remains a problem beyond
technical reach. Meanwhile, it is important to note that in contrast to the
general Kerr metric, the throat metric is not asymptotically flat. This
difference naturally shows itself in the potentials, too. In other words,
the extreme Kerr throat metric represents such a local region that the
geometry of interest is different than the geometry of the general Kerr
metric. The steeply rising potential prevents any particle (or field) flow
to infinity to make superradiance. Hence, we can mentally say that the
behavior of particles in two geometries must be considered separately.

In order to separate equations we employ the well-known method due to
Chandrasekhar so that we prefer to label the set of equations as
Chandrasekhar-Dirac (CD) equations. We separate the $y$ \ (axial) and $%
\theta $ (angular) dependence in such a way that the resulting \ axial
equation remains independent of mass. This leads us to an exact solution
irrespective of mass. The angular equation on the other hand depends
strictly on the mass. For the massless case (which we refer to as neutrino
equation) the angular equation reduces to a confluent Heun equation [8].
When the mass is nonzero, however, we can not identify our equations but
instead we express them as a set of linear equations suitable for numerical
analysis.

Organization of the paper is as follows: in section II we review the near
horizon geometry of an extreme Kerr BH and separation of variables of the CD
equation. Solution of the axial equation follows in section III. The
massless and massive cases are discussed in sections IV and V, respectively.
The reduction of our equations into one-dimensional Schr\"{o}dinger-type
equations with their conserved currents and superradiance are all included
in section VI. The paper ends with a conclusion in section VII.

\ 

\section{EXTREME KERR THROAT GEOMETRY AND SEPARATION OF DIRAC EQUATION ON IT}

The extreme Kerr metric in the Boyer-Lindquist coordinates is given by

\begin{equation}
ds^{2}=e^{2\nu }d\widetilde{t}^{2}-e^{2\psi }\left( d\widetilde{\phi }%
+\omega d\widetilde{t}\right) ^{2}-\rho ^{2}\left( \frac{d\widetilde{r}^{2}}{%
\widetilde{\Delta }}+d\theta ^{2}\right)  \eqnum{1}
\end{equation}

where

\[
e^{2\nu }=\frac{\widetilde{\Delta }\rho ^{2}}{\left( \widetilde{r}%
^{2}+M^{2}\right) ^{2}-\widetilde{\Delta }M^{2}\sin ^{2}\theta } 
\]

\[
e^{2\left( \nu +\psi \right) }=\widetilde{\Delta }\sin ^{2}\theta 
\]

\begin{equation}
\widetilde{\Delta }=\left( \widetilde{r}-M\right) ^{2}  \eqnum{2}
\end{equation}

\[
\omega =\frac{2M^{2}\widetilde{r}e^{2\nu }}{\widetilde{\Delta }\rho ^{2}} 
\]

\[
\rho ^{2}=\widetilde{r}^{2}+M^{2}\cos ^{2}\theta 
\]

In the extreme case, both the total mass $M$ and the rotation parameter $a$
become identical so that the angular momentum $J=M^{2}$ and the extremal
horizon corresponds to $\widetilde{r}=M$. The area of the horizon is $A=8\pi
J$.

To describe the near horizon (or throat) limit of the extreme Kerr metric,
due originally to Bardeen and Horowitz [7], one can set

\[
\widetilde{r}=M+\lambda r 
\]

\begin{equation}
\widetilde{t}=\frac{t%
{\acute{}}%
}{\lambda }  \eqnum{3}
\end{equation}

\[
\widetilde{\phi }=\phi -\frac{t%
{\acute{}}%
}{2\lambda M} 
\]

and takes the limit $\lambda \rightarrow 0.$ In these new coordinates, the
throat metric is obtained as

\begin{equation}
ds^{2}=F\left[ \frac{r^{2}}{r_{0}^{2}}dt%
{\acute{}}%
^{\text{ }2}-\frac{r_{0}^{2}}{r^{2}}dr^{2}-r_{0}^{2}d\theta ^{2}\right] -%
\frac{r_{0}^{2}\sin ^{2}\theta }{F}\left( d\phi +\frac{r}{r_{0}^{2}}dt%
{\acute{}}%
\right) ^{2}  \eqnum{4}
\end{equation}
\ 

where

\[
F=\frac{1+\cos ^{2}\theta }{2} 
\]

\begin{equation}
r_{0}^{2}=2M^{2}  \eqnum{5}
\end{equation}

We set further, for simplicity, $r_{0}^{2}=1$\ . This throat spacetime has
no longer asymptotic flatness.

Finally, passing to more general coordinates,

\[
y=\frac{1}{2r}\left[ r^{2}\left( 1+t%
{\acute{}}%
^{\text{ }2}\right) -1\right] 
\]

\begin{equation}
\cot t=\frac{1}{2t%
{\acute{}}%
r}\left[ r^{2}\left( 1-t%
{\acute{}}%
^{\text{ }2}\right) +1\right]  \eqnum{6}
\end{equation}

\[
\phi =\varphi +\ln \left| \frac{\cos t+y\sin t}{1+t%
{\acute{}}%
r}\right| 
\]

we can write the throat metric (4) as follows

\begin{equation}
ds^{2}=F\left[ \left( 1+y^{2}\right) dt^{2}-\frac{dy^{2}}{1+y^{2}}-d\theta
^{2}\right] -\frac{\sin ^{2}\theta }{F}\left( d\varphi +ydt\right) ^{2} 
\eqnum{7}
\end{equation}

The metric functions in (7) depend only on the variable $\theta $ thus as
expected in the search of solution for the Dirac equation the angular
equation forms the crux of the problem. The coordinates $-\infty <t<\infty $%
\ , $-\infty <y<\infty $ cover the entire, singularity free spacetime. The
Killing vector $\frac{\partial }{\partial t}$ is not timelike everywhere; it
admits a region (for $\sin ^{2}\theta >0.536$) in which it becomes
spacelike. Therefore by a coordinate transformation this particular region
is transformable to the spacetime of colliding plane waves [9]. Recently, it
has also been shown that the metric (7) can be obtained as a solution to
dilaton-axion gravity which is similar to the rotating BR spacetime [10].

The singularity free character can best be seen by checking the Weyl scalar $%
\Psi _{2}$ and the Kretschmann scalar:

\begin{equation}
\Psi _{2}=\frac{2}{\left( 1+\cos ^{2}\theta \right) ^{3}}\left[ 3\cos
^{2}\theta -1+i\cos \theta \left( \cos ^{2}\theta -3\right) \right] 
\eqnum{8}
\end{equation}

\begin{equation}
R_{\mu \nu \rho \sigma }R^{\mu \nu \rho \sigma }=\frac{192\sin ^{2}\theta }{%
\left( 1+\cos ^{2}\theta \right) ^{6}}\left[ \left( 1+\cos ^{2}\theta
\right) ^{2}-16\cos ^{2}\theta \right]  \eqnum{9}
\end{equation}

When the backreaction of the spin-$\frac{1}{2}$ test particles on the
background geometry is neglected, the Dirac field equation is given by the
CD equations [1] on a fixed space-time (7).

We choose a complex null tetrad $\left\{ l,n,m,\overline{m}\right\} $ that
satisfies the orthogonality conditions, $l.n=-m.\overline{m}=1.$ We note
that, throughout the paper, a bar over a quantity denotes complex
conjugation. Thus the covariant one-forms can be written as

\[
\sqrt{2}l=\sqrt{F(1+y^{2})}dt-\frac{\sqrt{F}}{\sqrt{1+y^{2}}}dy 
\]

\begin{equation}
\sqrt{2}n=\sqrt{F(1+y^{2})}dt+\frac{\sqrt{F}}{\sqrt{1+y^{2}}}dy  \eqnum{10}
\end{equation}
\[
\sqrt{2}m=\frac{iy\sin \theta }{\sqrt{F}}dt+\sqrt{F}d\theta +\frac{i\sin
\theta }{\sqrt{F}}d\varphi 
\]

\[
\sqrt{2}\overline{m}=-\frac{iy\sin \theta }{\sqrt{F}}dt+\sqrt{F}d\theta -%
\frac{i\sin \theta }{\sqrt{F}}d\varphi 
\]

and their corresponding directional derivatives are

\[
\sqrt{2}D=\frac{1}{\sqrt{F(1+y^{2})}}\partial _{t}+\frac{\sqrt{1+y^{2}}}{%
\sqrt{F}}\partial _{y} 
\]

\[
\sqrt{2}\Delta =\frac{1}{\sqrt{F(1+y^{2})}}\partial _{t}-\frac{\sqrt{1+y^{2}}%
}{\sqrt{F}}\partial _{y} 
\]

\begin{equation}
\sqrt{2}\delta =-\frac{1}{\sqrt{F}}\partial _{\theta }-\frac{i\sqrt{F}}{\sin
\theta }\partial _{\varphi }  \eqnum{11}
\end{equation}

\[
\sqrt{2}\overline{\delta }=-\frac{1}{\sqrt{F}}\partial _{\theta }+\frac{i%
\sqrt{F}}{\sin \theta }\partial _{\varphi } 
\]

One can determine the nonzero NP complex spin coefficients [11] in the above
null-tetrad as

\[
\pi =-\tau =\frac{\sin \theta \left( \cos \theta -i\right) }{\left(
2F\right) ^{\frac{3}{2}}} 
\]

\begin{equation}
\varepsilon =\gamma =\frac{y}{2\sqrt{2F(1+y^{2})}}  \eqnum{12}
\end{equation}

\[
\alpha =-\beta =\frac{2\cot \theta -i\sin \theta }{2\left( 2F\right) ^{\frac{%
3}{2}}} 
\]

The CD equations in the NP formalism are then[1]

\[
\left( D+\varepsilon -\rho \right) F_{1}+\left( \overline{\delta }+\pi
-\alpha \right) F_{2}=i\mu _{p}G_{1} 
\]

\[
\left( \Delta +\mu -\gamma \right) F_{2}+\left( \delta +\beta -\tau \right)
F_{1}=i\mu _{p}G_{2} 
\]

\begin{equation}
\left( D+\overline{\varepsilon }-\overline{\rho }\right) G_{2}-\left( \delta
+\overline{\pi }-\overline{\alpha }\right) G_{1}=i\mu _{p}F_{2}  \eqnum{13}
\end{equation}

\[
\left( \Delta +\overline{\mu }-\overline{\gamma }\right) G_{1}-\left( 
\overline{\delta }+\overline{\beta }-\overline{\tau }\right) G_{2}=i\mu
_{p}F_{1} 
\]

where $\mu ^{\ast }=\sqrt{2}\mu _{p}$ is the mass of the Dirac particle.

The form of the CD equations suggests that we introduce [1,6]

\[
F_{1}=f_{1}\left( y\right) A_{1}\left( \theta \right) e^{i\left( \sigma
t+m\varphi \right) } 
\]

\[
G_{1}=g_{1}\left( y\right) A_{2}\left( \theta \right) e^{i\left( \sigma
t+m\varphi \right) } 
\]

\begin{equation}
F_{2}=f_{2}\left( y\right) A_{3}\left( \theta \right) e^{i\left( \sigma
t+m\varphi \right) }  \eqnum{14}
\end{equation}

\[
G_{2}=g_{2}\left( y\right) A_{4}\left( \theta \right) e^{i\left( \sigma
t+m\varphi \right) } 
\]

\bigskip where $\sigma $ is the frequency of the corresponding Compton wave
of the Dirac particle and $m$ is the azimuthal quantum number of the wave.
Our convention is that $\sigma $ is always positive.

Inserting for the appropriate spin coefficients (12) with the spinors (14)
into the four coupled CD equations (13), we obtain

\[
\frac{\left( \widetilde{Z}f_{1}\right) }{f_{2}}=i\mu ^{\ast }\frac{g_{1}}{%
f_{2}}\frac{A_{2}}{A_{1}}\sqrt{F}-\frac{\left( LA_{3}\right) }{A_{1}} 
\]

\[
\frac{\left( \overline{\widetilde{Z}}f_{2}\right) }{f_{1}}=-i\mu ^{\ast }%
\frac{g_{2}}{f_{1}}\frac{A_{4}}{A_{3}}\sqrt{F}-\frac{\left(
L^{+}A_{1}\right) }{A_{3}} 
\]

\begin{equation}
\frac{\left( \widetilde{Z}g_{2}\right) }{g_{1}}=i\mu ^{\ast }\frac{f_{2}}{%
g_{1}}\frac{A_{3}}{A_{4}}\sqrt{F}-\frac{\left( \pounds ^{+}A_{2}\right) }{%
A_{4}}  \eqnum{15}
\end{equation}

\[
-\frac{\left( \overline{\widetilde{Z}}g_{1}\right) }{g_{2}}=i\mu ^{\ast }%
\frac{f_{1}}{g_{2}}\frac{A_{1}}{A_{2}}\sqrt{F}-\frac{\left( \pounds
A_{4}\right) }{A_{2}} 
\]

where the axial and the angular operators are

\[
\widetilde{Z}=\sqrt{1+y^{2}}\partial _{y}+\frac{1}{2\sqrt{1+y^{2}}}\left[
y+2i\sigma \right] 
\]

\begin{equation}
\overline{\widetilde{Z}}=\sqrt{1+y^{2}}\partial _{y}+\frac{1}{2\sqrt{1+y^{2}}%
}\left[ y-2i\sigma \right]  \eqnum{16}
\end{equation}

and

\[
L=\partial _{\theta }+\frac{mF}{\sin \theta }+\frac{1}{2F}\left( \frac{\cos
^{3}\theta }{\sin \theta }+\frac{i\sin \theta }{2}\right) \text{ } 
\]

\[
L^{+}=\partial _{\theta }-\frac{mF}{\sin \theta }+\frac{1}{2F}\left( \frac{%
\cos ^{3}\theta }{\sin \theta }+\frac{i\sin \theta }{2}\right) 
\]

\begin{equation}
\text{\pounds }=\partial _{\theta }+\frac{mF}{\sin \theta }+\frac{1}{2F}%
\left( \frac{\cos ^{3}\theta }{\sin \theta }-\frac{i\sin \theta }{2}\right) 
\eqnum{17}
\end{equation}

\[
\text{\pounds }^{+}=\partial _{\theta }-\frac{mF}{\sin \theta }+\frac{1}{2F}%
\left( \frac{\cos ^{3}\theta }{\sin \theta }-\frac{i\sin \theta }{2}\right) 
\]

respectively. One can easily see that $L=\overline{\pounds }$ and $L^{+}=$ $%
\overline{\pounds ^{+}}$.

Further, choosing $f_{1}=g_{2}$, $f_{2}=g_{1}$, $A_{1}=\overline{A_{2}}$ and 
$A_{3}=\overline{A_{4}}$ and introducing the separation constant as $%
i\lambda $,\ where $\lambda $ is a real constant, we can separate Dirac
equation (15) into axial and angular parts

\begin{equation}
\overline{\widetilde{Z}}g_{1}=-i\lambda g_{2}  \eqnum{18}
\end{equation}

\begin{equation}
\widetilde{Z}g_{2}=i\lambda g_{1}  \eqnum{19}
\end{equation}

and

\begin{equation}
LA_{3}+i\mu ^{\ast }A_{2}\sqrt{F}=i\lambda A_{1}  \eqnum{20}
\end{equation}

\begin{equation}
L^{+}A_{1}+i\mu ^{\ast }A_{4}\sqrt{F}=i\lambda A_{3}  \eqnum{21}
\end{equation}

It is clear from equations (18,19) that $g_{1}=\overline{g}_{2}.$

\section{SOLUTION OF THE\ AXIAL EQUATION}

\bigskip If we decouple the axial equations (18,19) in equation (18) to get $%
g_{1}$, we obtain

\begin{equation}
\widetilde{Z}(\overline{\widetilde{Z}}g_{1})=\lambda ^{2}g_{1}  \eqnum{22}
\end{equation}

Similarly one can decouple the axial equations in equation (18) for $g_{2}.$
The explicit form of equation (22) can be obtained as

\begin{equation}
\left( 1+y^{2}\right) g_{1}^{\prime \prime }(y)+2yg_{1}^{\prime }(y)+\frac{1%
}{1+y^{2}}\left( \frac{1}{2}+\sigma ^{2}+\frac{y^{2}}{4}-\lambda ^{2}\left(
1+y^{2}\right) +i\sigma y\right) g_{1}(y)=0  \eqnum{23}
\end{equation}

( Throughout the paper, a prime denotes the derivative with respect to its
argument.)

Thus the solutions of the decoupled equations for $g_{1}$, equation (23),
and $g_{2}$ (not given here) can be found in terms of the associated
Legendre functions as follows

\[
g_{1}(y)=c_{1}P_{\lambda -\frac{1}{2}}^{\widehat{\beta }}\left( iy\right) 
\]

\begin{equation}
g_{2}(y)=c_{2}P_{\lambda -\frac{1}{2}}^{\widehat{\beta }}\left( -iy\right)  
\eqnum{24}
\end{equation}

where

\begin{equation}
\widehat{\beta }=\sqrt{\sigma ^{2}+\frac{\sigma }{2}+\frac{1}{4}}  \eqnum{25}
\end{equation}

and $c_{1},$ $c_{2}$ are complex constants.

\bigskip Here, due to the physical necessities, we considered only the first
kind of the associated Legendre functions. Although solutions (24) seem like
complex solutions, it is possible to draw the real functions from the above
associated Legendre functions. We may define

\begin{equation}
\lambda =\widetilde{m}+\frac{1}{2}\text{ \ with \ }\widetilde{m}=1,2,3.... 
\eqnum{26}
\end{equation}

\bigskip and

\begin{equation}
\sigma =\frac{1}{4}\left( \sqrt{16\tilde{n}^{2}-3}-1\right)  \eqnum{27}
\end{equation}

so that

\begin{equation}
\widehat{\beta }=\tilde{n}\text{ \ \ \ \ \ with \ \ \ }\tilde{n}=-\widetilde{%
m},-\widetilde{m}+1,.....-1,1,......\widetilde{m}-1,\widetilde{m}\text{ } 
\eqnum{28}
\end{equation}

In order to get the real functions for solutions (22), the required
condition is: $\widetilde{m}-\left| \tilde{n}\right| =$even number.

It is worth also drawing attention to the following remarks:

{\it i)} In the case of $\lambda =0,$ equations (18,19) reduce to simple
first order differential equations which admit the following solutions

\[
g_{1}(y)=c_{3}\left( 1+y^{2}\right) ^{-\frac{1}{4}}e^{i\sigma \tan ^{-1}(y)} 
\]

\begin{equation}
g_{2}(y)=c_{4}\left( 1+y^{2}\right) ^{-\frac{1}{4}}e^{-i\sigma \tan
^{-1}\left( y\right) }  \eqnum{29}
\end{equation}

with $c_{3},c_{4}$ complex constants.

These two solutions can be interpreted as representing ingoing and outgoing
waves.

{\it ii)} In the case of $\lambda =\frac{1}{2},$ we obtain the following
complex solutions from equations (18,19)

\[
g_{1}(y)=c_{5}\left( \frac{iy+1}{iy-1}\right) ^{\frac{\widehat{\beta }}{2}%
}+c_{6}\left( \frac{iy+1}{iy-1}\right) ^{-\frac{\widehat{\beta }}{2}} 
\]

\begin{equation}
g_{2}(y)=c_{7}\left( \frac{1-iy}{1+iy}\right) ^{\frac{\widehat{\beta }}{2}%
}+c_{8}\left( \frac{1-iy}{1+iy}\right) ^{-\frac{\widehat{\beta }}{2}} 
\eqnum{30}
\end{equation}

where again $c_{j}$ with $j=5,6,7,8$ are complex constants.

\section{REDUCTION OF THE ANGULAR EQUATION TO HEUN EQUATION: THE MASSLESS
CASE}

In this section, we shall show that the angular equations (20,21) for the
neutrino particles can be decoupled to the confluent Heun equation. To the
end that let us reconsider equations (20,21) in the explicit form for $\mu
^{\ast }=0,$

\begin{equation}
A_{3}^{\prime }(\theta )+(K+G)A_{3}(\theta )=i\lambda A_{1}(\theta ) 
\eqnum{31}
\end{equation}

\begin{equation}
A_{1}^{\prime }(\theta )+(K-G)A_{1}(\theta )=i\lambda A_{3}(\theta ) 
\eqnum{32}
\end{equation}

where

\begin{equation}
K=\frac{1}{2F}\left( \frac{\cos ^{3}\theta }{\sin \theta }+\frac{i\sin
\theta }{2}\right)  \eqnum{33}
\end{equation}

\begin{equation}
G=\frac{mF}{\sin \theta }  \eqnum{34}
\end{equation}

By introducing the scalings

\begin{equation}
A_{1}(\theta )=H_{1}(\theta )e^{-\int \left( K-G\right) d\theta }  \eqnum{35}
\end{equation}

\begin{equation}
A_{3}(\theta )=H_{3}(\theta )e^{-\int \left( K+G\right) d\theta }  \eqnum{36}
\end{equation}

one gets

\begin{equation}
H_{1}^{\prime }(\theta )=i\lambda H_{3}(\theta )e^{-\int 2Gd\theta } 
\eqnum{37}
\end{equation}

\begin{equation}
H_{3}^{\prime }(\theta )=i\lambda H_{1}(\theta )e^{\int 2Gd\theta } 
\eqnum{38}
\end{equation}

If we decouple equations (37,38) in equation (37) for $H_{1}(\theta )$, we
get

\begin{equation}
H_{1}^{\prime \prime }(\theta )+2GH_{1}^{\prime }(\theta )+\lambda
^{2}H_{1}(\theta )=0  \eqnum{39}
\end{equation}

In similar fashion, we find, for $H_{3}(\theta ),$

\begin{equation}
H_{3}^{\prime \prime }(\theta )-2GH_{3}^{\prime }(\theta )+\lambda
^{2}H_{3}(\theta )=0  \eqnum{40}
\end{equation}

Introducing a new variable $\theta =\cos ^{-1}\left( 1-2z\right) ,$
equations (39,40) turn out to be

\begin{equation}
H_{1}^{\prime \prime }(z)+\left( -2m+\frac{\frac{1}{2}+m}{z}+\frac{\frac{1}{2%
}-m}{z-1}\right) H_{1}^{\prime }(z)-\frac{\lambda ^{2}}{z(z-1)}H_{1}(z)=0 
\eqnum{41}
\end{equation}

\begin{equation}
H_{3}^{\prime \prime }(z)+\left( -2m+\frac{m-\frac{1}{2}}{z}+\frac{m+\frac{1%
}{2}}{z-1}\right) H_{3}^{\prime }(z)-\frac{\lambda ^{2}}{z(z-1)}H_{3}(z)=0 
\eqnum{42}
\end{equation}

Let us recall the general confluent form of Heun equation [8],

\begin{equation}
H^{\prime \prime }(z)+\left( A+\frac{B}{z}+\frac{C}{z-1}\right) H^{\prime
}(z)-\frac{DBz-h}{z(z-1)}H(z)=0  \eqnum{43}
\end{equation}

Drawing the similarities between equation (43) and equations (41,42), we
observe the following correspondences:

{\it a) }For equation (41),

\begin{equation}
D=0,\text{ }h=\lambda ^{2},\text{ }A=-2m,\text{ }B=\frac{1}{2}+m\text{ and }%
C=\frac{1}{2}-m  \eqnum{44}
\end{equation}

{\it b) }For equation (42),

\begin{equation}
D=0,\text{ }h=\lambda ^{2},\text{ }A=-2m,\text{ }B=m-\frac{1}{2}\text{ and }%
C=m+\frac{1}{2}  \eqnum{45}
\end{equation}

The confluent Heun equation (43), with its accessory parameter $h$, has two
regular singular points at $z=0,1$ with exponents $(0,1-B)$ and $(0,1-C)$,
respectively, \ as well as an irregular singularity at the infinity. In the
vicinity of the point $z=0$, its power series can be written as

\begin{equation}
H(D,A,B,C,h;z)=\sum_{j=0}^{\infty }W_{j}z^{j}  \eqnum{46}
\end{equation}

and the coefficient $W_{j}$ satisfies a three-term recurrence relation [8],

\begin{eqnarray}
W_{0} &=&1,\text{ \ \ \ \ \ \ \ }W_{1}=\frac{-h}{B}  \eqnum{47} \\
\left( j+1\right) \left( j+B\right) W_{j+1}-A(j-1+D)W_{j-1} &=&\left[
j\left( j-1-A+B+C\right) -h\right] W_{j}  \nonumber
\end{eqnarray}

It is also possible to obtain the power series solution in the vicinity of
the point $z=1$ by a linear transformation interchanging the regular
singular points $z=0$ and $z=1$. Namely, $z\rightarrow 1-z$.

Expansion of solutions to the confluent Heun equation in terms of the
hypergeometric and confluent hypergeometric function can be seen in [8]. In
Ref.[8], it is also shown that the confluent Heun equation can be normalized
to constitute a group of orthogonal complete functions and the confluent
Heun equation also admits quasipolynomial solutions for particular values of
the parameters.

Since $D=0$ in our case, it follows from the three-term recurrence relation
that $H(D,A,B,C,h;z)$ is a polynomial solution if $W_{1}(h)=0$, where $W_{1}$
stands for a polynomial of degree $1$ in $h$. Namely , there is only one
eigenvalue $h_{i}$ for $h$ such that $W_{1}(h_{i})=0$ (i.e. $\lambda =0$).

\section{REDUCTION OF THE ANGULAR EQUATION INTO A SET OF LINEAR FIRST ORDER
DIFFERENTIAL EQUATIONS: THE CASE WITH MASS}

To complete our analysis of the angular equation, we need to discuss the
angular equation for the Dirac particles with mass.

The angular equations (20,21) can be rewritten in the following forms

\begin{equation}
LA_{3}+i\mu _{p}\sqrt{1+\cos ^{2}\theta }\overline{A_{1}}=i\lambda A_{1} 
\eqnum{48}
\end{equation}

\begin{equation}
L^{+}A_{1}+i\mu _{p}\sqrt{1+\cos ^{2}\theta }\overline{A_{3}}=i\lambda A_{3}
\eqnum{49}
\end{equation}

With substitutions

\begin{equation}
A_{1}(\theta )=\left( A_{0}(\theta )+iB_{0}\left( \theta \right) \right)
e^{\int \frac{\cos ^{3}\theta }{2\sin \theta F}d\theta }  \eqnum{50}
\end{equation}

\begin{equation}
A_{3}(\theta )=\left( M_{0}(\theta )+iN_{0}\left( \theta \right) \right)
e^{\int \frac{\cos ^{3}\theta }{2\sin \theta F}d\theta }  \eqnum{51}
\end{equation}

we can transform equations (48,49) into a set of first order differential
equations

\[
M_{0}^{\prime }(\theta )+GM_{0}(\theta )-\frac{\sin \theta }{4F}N_{0}\left(
\theta \right) =-\left( \lambda +\mu _{p}\sqrt{1+\cos ^{2}\theta }\right)
B_{0}\left( \theta \right) 
\]

\[
N_{0}^{\prime }(\theta )+GN_{0}(\theta )+\frac{\sin \theta }{4F}M_{0}\left(
\theta \right) =\left( \lambda -\mu _{p}\sqrt{1+\cos ^{2}\theta }\right)
A_{0}\left( \theta \right) 
\]

\begin{equation}
A_{0}^{\prime }(\theta )-GA_{0}(\theta )-\frac{\sin \theta }{4F}B_{0}\left(
\theta \right) =-\left( \lambda +\mu _{p}\sqrt{1+\cos ^{2}\theta }\right)
N_{0}\left( \theta \right)  \eqnum{52}
\end{equation}

\[
B_{0}^{\prime }(\theta )-GB_{0}(\theta )+\frac{\sin \theta }{4F}A_{0}\left(
\theta \right) =\left( \lambda -\mu _{p}\sqrt{1+\cos ^{2}\theta }\right)
M_{0}\left( \theta \right) 
\]

Introducing a new variable $x=\cos \theta $ and with the further
substitutions

\[
M_{0}(\theta )=\frac{1}{2}\left( m_{0}(\theta )+a_{0}(\theta )\right) \text{
\ \ \ \ \ \ \ \ \ \ \ \ \ \ \ \ \ \ \ \ \ \ \ }N_{0}(\theta )=\frac{1}{2}%
\left( n_{0}(\theta )+b_{0}(\theta )\right) \text{\ \ \ } 
\]

\begin{equation}
A_{0}(\theta )=\frac{1}{2}\left( m_{0}(\theta )-a_{0}(\theta )\right) \text{
\ \ \ \ \ \ \ \ \ \ \ \ \ \ \ \ \ \ \ \ \ \ }B_{0}(\theta )=\frac{1}{2}%
\left( n_{0}(\theta )-b_{0}(\theta )\right)  \eqnum{53}
\end{equation}

we may obtain the final form of the set as linear first order differential
equations

\[
m_{0}^{\prime }(x)+\alpha _{1}a_{0}(x)+\left( \alpha _{2}-\alpha _{3}\right)
n_{0}(x)=0 
\]

\[
a_{0}^{\prime }(x)+\alpha _{1}m_{0}(x)+\left( \alpha _{4}+\alpha _{3}\right)
b_{0}(x)=0 
\]

\begin{equation}
n_{0}^{\prime }(x)+\alpha _{1}b_{0}(x)-\left( \alpha _{2}+\alpha _{3}\right)
m_{0}(x)=0  \eqnum{54}
\end{equation}

\[
b_{0}^{\prime }(x)+\alpha _{1}n_{0}(x)-\left( \alpha _{4}-\alpha _{3}\right)
a_{0}(x)=0 
\]

where

\[
\alpha _{1}=-\frac{m\left( 1+x^{2}\right) }{2\left( 1-x^{2}\right) }\text{ \
\ \ \ \ \ \ \ \ \ \ \ \ \ \ \ \ \ \ \ \ \ \ }\alpha _{2}=\frac{1}{2\left(
1+x^{2}\right) }-\frac{\lambda }{\sqrt{1-x^{2}}} 
\]

\begin{equation}
\alpha _{3}=\frac{\mu _{p}\sqrt{1+x^{2}}}{\sqrt{1-x^{2}}}\text{ \ \ \ \ \ \
\ \ \ \ \ \ \ \ \ \ \ \ \ \ \ \ }\alpha _{2}=\frac{1}{2\left( 1+x^{2}\right) 
}+\frac{\lambda }{\sqrt{1-x^{2}}}  \eqnum{55}
\end{equation}

Although system (54) does not seem to\ be solved analytically, one may
develop an appropriate numerical technique to study it. In the literature,
there may exist such interesting systems which are more or less of this type.

\section{ THE REDUCTION OF DIRAC EQUATION TO ONE-DIMENSIONAL \
SCHR\"{O}DINGER-TYPE EQUATION WITH A CONSERVED CURRENT}

\bigskip It is possible to get more compact forms the axial equations
(18,19) by introducing the scalings

\begin{equation}
g_{1}(y)=Z_{1}(y)\left( 1+y^{2}\right) ^{-\frac{1}{4}}  \eqnum{56}
\end{equation}

\begin{equation}
g_{2}(y)=Z_{2}(y)\left( 1+y^{2}\right) ^{-\frac{1}{4}}  \eqnum{57}
\end{equation}

and applying the coordinate transformation $y=\tan u$ , the axial equations
take the forms

\begin{equation}
Z_{1}^{\prime }(u)-i\sigma Z_{1}(u)=-i\lambda XZ_{2}(u)  \eqnum{58}
\end{equation}

\begin{equation}
Z_{2}^{\prime }(u)+i\sigma Z_{2}(u)=i\lambda XZ_{1}(u)  \eqnum{59}
\end{equation}

\bigskip where \ $X=\frac{1}{\sqrt{1+y^{2}}}\equiv \cos u.$

Letting

\begin{equation}
Z_{1}(u)=\frac{iP_{1}(u)-P_{2}(u)}{2}  \eqnum{60}
\end{equation}

\begin{equation}
Z_{2}(u)=\frac{iP_{1}(u)+P_{2}(u)}{2}  \eqnum{61}
\end{equation}

we can combine equations (58,59) to give

\begin{equation}
P_{1}^{\prime }(u)=-E_{+}P_{2}(u)  \eqnum{62}
\end{equation}

\begin{equation}
P_{2}^{\prime }(u)=E_{-}P_{1}(u)  \eqnum{63}
\end{equation}

where

\begin{equation}
E_{+}=\sigma +\lambda X  \eqnum{64}
\end{equation}

\begin{equation}
E_{-}=\sigma -\lambda X  \eqnum{65}
\end{equation}

Decoupling is attained, by introducing

\begin{equation}
P_{1}(u)=\sqrt{E_{+}}T(u)  \eqnum{66}
\end{equation}

\begin{equation}
P_{2}(u)=\sqrt{E_{-}}S(u)  \eqnum{67}
\end{equation}

where we obtain a pair of one-dimensional Schr\"{o}dinger-type equations

\begin{equation}
T^{\prime \prime }(u)+V_{1}T(u)=0  \eqnum{68}
\end{equation}

\begin{equation}
S^{\prime \prime }(u)+V_{2}S(u)=0  \eqnum{69}
\end{equation}

with the potentials

\begin{equation}
V_{1}=\sigma ^{2}-\lambda ^{2}X^{2}(1+\frac{y^{2}}{4E_{+}^{2}})+\frac{%
\lambda X}{2E_{+}^{2}}\left( \sigma \left( 1+2y^{2}\right) +\lambda
X^{3}\right)  \eqnum{70}
\end{equation}

\begin{equation}
V_{2}=\sigma ^{2}-\lambda ^{2}X^{2}(1+\frac{y^{2}}{4E_{-}^{2}})-\frac{%
\lambda X}{2E_{-}^{2}}\left( \sigma \left( 1+2y^{2}\right) -\lambda
X^{3}\right)  \eqnum{71}
\end{equation}

One can easily observe that for $y\rightarrow \pm \infty $ the potentials
diverge. This result stems from the fact that our space-time is not
asymptotically flat.

To examine the existence of the superradiance, one may consider the
conserved net current of Dirac particles [1]. In other words, the rate $%
\left( \frac{\partial N}{\partial t}\right) _{in}$ at which particles
falling through the horizon per unit time, which must be negative for the
superradiance to occur

\begin{equation}
\left( \frac{\partial N}{\partial t}\right) _{in}=-\left( \int \sqrt{-g}%
J^{y}d\theta d\varphi \right) \left| _{Horizon}\right. <0  \eqnum{72}
\end{equation}

where $g$ is the determinant of the spacetime metric and $J^{y}$ is the
axial component of the neutrino particle current.We recall from metric (7)
that we have

\begin{equation}
\sqrt{-g}=F\sin \theta  \eqnum{73}
\end{equation}

It is clear from transformations (3,6) that the horizon of metric (7)
corresponds to $y\rightarrow \left( -\infty \right) $. In other words,
integral (72) is taken over $y\rightarrow \left( -\infty \right) $.

In the more standard spinor formalism, $J^{y}$ is introduced as [1]

\begin{equation}
\frac{1}{\sqrt{2}}J^{y}=\sigma _{AB^{\prime }}^{y}\left( P^{A}\overline{P}%
^{B^{\prime }}+Q^{A}\overline{Q}^{B^{\prime }}\right)  \eqnum{74}
\end{equation}

\bigskip where

\begin{equation}
\sigma _{AB^{\prime }}^{y}=\frac{1}{\sqrt{2}}\left( 
\begin{array}{cc}
\sqrt{\frac{1+y^{2}}{F}} & 0 \\ 
0 & -\sqrt{\frac{1+y^{2}}{F}}
\end{array}
\right)  \eqnum{75}
\end{equation}

In this notation, the basic spinors defined by $P^{A}$ and $\overline{Q}%
^{A^{^{\prime }}}$ correspond to [6],

\begin{eqnarray}
P^{0} &=&F_{1},\text{ \ \ \ \ \ \ \ \ \ \ }P^{1}=F_{2}  \eqnum{76} \\
\overline{Q}^{0^{\prime }} &=&-G2,\text{ \ \ \ \ \ \ \ }\overline{Q}%
^{1^{\prime }}=G1  \nonumber
\end{eqnarray}

\bigskip We evaluate $J^{y}$ as

\begin{equation}
J^{y}=\sqrt{\frac{1+y^{2}}{F}}\left( \left| g_{2}\right| ^{2}-\left|
g_{1}\right| ^{2}\right) \left( \left| A_{1}\right| ^{2}+\left| A_{3}\right|
^{2}\right)  \eqnum{77}
\end{equation}

Assuming that the angular functions $A_{1}\left( \theta \right) $ and $%
A_{3}\left( \theta \right) $ are normalized to unity, the integral in
equation (72) yields that

\begin{equation}
\int \sqrt{-g}J^{y}d\theta d\varphi =3.246\pi \left( \left| Z_{2}\right|
^{2}-\left| Z_{1}\right| ^{2}\right)  \eqnum{78}
\end{equation}

From equations (60), (61), (62) and (63), we successively find

\begin{eqnarray}
\left| Z_{2}\right| ^{2}-\left| Z_{1}\right| ^{2} &=&\frac{i}{2}\left( P_{1}%
\overline{P}_{2}-P_{2}\overline{P}_{1}\right)  \eqnum{79} \\
&=&\frac{-i}{2E_{+}}\left[ P_{1},\overline{P}_{1}\right] _{u}  \nonumber
\end{eqnarray}

where $\left[ P_{1},\overline{P}_{1}\right] _{u}$ is the Wronskian.

Therefore, in order to check the existence of the superradiance, it will
suffice to seek the solution for $P_{1}$ at the horizon.

The reality that the potentials $V_{1}$ and $V_{2}$ become infinite both at
the horizon and $y\rightarrow \infty $,\ leads us to think the problem as a
problem of particle in the infinite potential well. Since the particles are
bound inside the well, the principal physical fact requires that the
solutions of the wave equations (68,69) must be identically zero at the
walls (the horizon and $y\rightarrow \infty $). Clearly, the Wronskian
vanishes at the horizon and it follows that the number of particles exiting
the horizon per unit time is zero.\ Consequently, similar to the general
Kerr background [1,3], there is also no superradiance in the extreme Kerr
throat geometry.

\section{CONCLUSION}

Our aim in this paper was to do more than separating the Dirac equation in a
sector of Kerr, namely the extremal Kerr throat geometry and obtain exact
solutions if possible. This premise has mostly been accomplished and they
definitely will contribute to the wave mechanical aspects of \ spin-$\frac{1%
}{2}$ particles prior infalling into the extreme Kerr BH.

In the general Kerr background the radial Dirac equation was the harder part
to be tackled compared with the angular part [1]. In the present problem of
the extremal Kerr throat we have the opposite case: the axial part poses no
more difficulty than the angular part does. For the massless case, we
overcome the difficulty and attain exact solution in terms of Heun
polynomials. Inclusion of mass prevents this reduction and as a result we
are unable to express the angular equation in terms of a set of known
equations. This part of the problem can be handled numerically.
Alternatively, the angular equation casts into a pair of Schr\"{o}%
dinger-type equations. Unlike the scalar field case Dirac fields exhibit no
superradiance. The charge coupling of a Dirac particle to an extremal
Kerr-Newman BH in its near horizon limit may reveal more information
compared to the present case. This is the next stage of study that interests
us.

\end{document}